\documentclass[11pt]{article}
\usepackage{amssymb,amsmath,amsfonts}
\usepackage{graphicx}
\usepackage{graphics}
\usepackage{eepic,epsfig}

\begin{document}

\title{Fermionic vacuum polarization by a composite topological defect in
higher-dimensional space-time}
\author{E. R. Bezerra de Mello$^{1}$\thanks{%
E-mail: emello@fisica.ufpb.br}\, and A. A. Saharian$^{1,2}$\thanks{%
E-mail: saharian@ictp.it} \\
\\
\textit{$^1$Departamento de F\'{\i}sica-CCEN, Universidade Federal da Para%
\'{\i}ba}\\
\textit{58.059-970, Caixa Postal 5.008, Jo\~{a}o Pessoa, PB, Brazil}\vspace{%
0.3cm}\\
\textit{$^2$Department of Physics, Yerevan State University,}\\
\textit{1 Alex Manoogian Street, 0025 Yerevan, Armenia}}
\maketitle

\begin{abstract}
We investigate the vacuum polarization effects associated with a charged
massless spin-1/2 field in a higher-dimensional space-time, induced by a
composite topological defect. The defect is constituted by a global monopole
living on a three-brane and two-dimensional conical space transverse to the
latter. In addition, we assume the presence of an extra magnetic flux along
the core of the conical space. The heat kernel and the Feynman Green
function are presented in the form of a sum of two terms. The first one
corresponds to the contribution coming from the bulk with global monopole in
the absence of conical structure of the orthogonal two-space, and the second
one is induced by this structure and the magnetic flux. We explicitly
evaluate the part in the vacuum expectation value of the energy-momentum
tensor induced by the flux carrying conical structure. As in pure cosmic
string geometries, only the fractional part of the ratio of the magnetic
flux to flux quantum leads to non-trivial effects. The vacuum
energy-momentum tensor is an even function of this parameter. We show that
for strong gravitational fields corresponding to large values of the solid
angle deficit, the effects induced by the conical structure and flux are
exponentially suppressed.
\end{abstract}

\bigskip


\bigskip

\section{Introduction}

Symmetry braking phase transitions in the early universe have several
cosmological consequences and provide an important link between particle
physics and cosmology. In particular, within the framework of grand unified
theories various types of topological defects are predicted to be formed
\cite{Vile94}. Among them the cosmic strings are of special interest.
Although the recent observational data on the cosmic microwave background
radiation have excluded cosmic strings as seeds for structure formation,
they are still candidates for the generation of a number of interesting
physical effects such as the generation of gravitational waves and gamma ray
bursts. Recently, cosmic strings attract a renewed interest partly because a
variant of their formation mechanism is proposed in the framework of brane
inflation \cite{Sara02}. The non-trivial space-time around a cosmic string
leads to interesting effects in quantum field theory. The vacuum
polarization associated with scalar and fermionic fields in the geometry of
an idealized cosmic string, have been analyzed in \cite{Line87,Sahni} and
\cite{Frol87}, respectively. Moreover, considering the presence of the
magnetic flux along the cosmic string, there appears an additional
contribution to the vacuum polarization effect associated with charged
fields \cite{Dowk87}. The combined effects of the topology and boundaries in
the geometry of a cosmic string are investigated in \cite{Brev95,Mello3}.

Though the topological defects have been first analyzed in four-dimensional
space-time, they have been considered within the framework of the braneworld
scenario as well. By this scenario our four-dimensional world emerges as a
defect in a higher-dimensional space-time (for a review see \cite{Ruba01}).
Braneworlds naturally appear in the string/M-theory context and provide a
novel setting for discussing phenomenological and cosmological issues
related to extra dimensions. The models introduced by Randall and Sundrum
are particularly attractive \cite{RS}. The corresponding space-time contains
one or two Ricci-flat branes embedded on a five-dimensional anti-de Sitter
bulk. More recently, the Randall-Sundrum scenario is generalized to the case
of two extra dimensions by using a global string \cite{Cohen}. For the case
with three extra dimensions, magnetic monopole and global monopole have been
analyzed in \cite{Roessl} and \cite{Ola,Cho2}, respectively. In particular,
in \cite{Cho2} the authors have obtained the solution to the Einstein
equations considering a general $p$-dimensional Minkowski brane worldsheet
and a $d$-dimensional global monopole, with $d\geqslant 3$, in the
transverse extra dimensions with the core on the brane.

The investigation of quantum effects in corresponding braneworld models is
of considerable phenomenological interest, both in particle physics and
cosmology. Quantum effects provide a natural alternative for the
stabilization of the radion fields. The corresponding vacuum energies give
contribution to both the brane and bulk cosmological constant and, hence,
has to be taken into account in the self-consistent formulation of the
braneworld dynamics. In recent papers we have investigated the vacuum
polarization effects associated with massless scalar \cite{Mello} and
fermionic \cite{Mello1} fields, respectively, in higher dimensional global
monopole space-time, in braneworld context. One-loop quantum effects of a
scalar field induced by a composite topological defect consisting a cosmic
string on a $p$-dimensional brane and a $(m+1)$-dimensional global monopole
in the transverse extra dimensions are considered in \cite{Mello2}. In order
to develop these analysis we have explicitly constructed the corresponding
Green functions. The objective of this paper is to complete the analysis,
studying the fermionic vacuum polarization effects associated with quantum
charged field propagating on six-dimensional space-time produced by a
composite defect, along the same line of investigation developed in \cite%
{Mello2}. Here we shall consider the monopole on the three-brane and the
cosmic string on the two-dimensional transverse space. In this analysis we
also allow the presence of an extra magnetic field, which can be understood
as producing a magnetic flux running along the core of the cosmic string
(about vacuum polarization energies for flux tube configurations see, for
instance, \cite{Bord99} and references therein). Note that fluxes by gauge
fields play an important role in higher-dimensional models including the
braneworld scenario (see, for instance, \cite{Doug07}). In particular, they
provide an stabilization mechanism for all moduli fields appearing in
various string compactifications. The problem under consideration is also of
separate interest as an example with gravitational and topology-induced
polarizations of the vacuum, where all calculations can be performed in a
closed form.

This paper is organized as follows. In section \ref{sec:heat}, we introduce
the structure of the space-time produced by the topological defect under
consideration. With the objective to construct the fermionic propagator in
this manifold, we first evaluate the heat kernel associated with the square
of the corresponding Dirac operator. Although the general expression for the
heat kernel is given in terms of series involving the modified Bessel
functions, a much simpler expression is obtained for a particular choice of
the parameters which codify the conical structure of the transverse
two-space and the fractional part of the ratio of the magnetic flux by the
quantum one. This special case is considered in section \ref{sec:special}.
In this section we also analyze the fermionic propagator in the coincidence
limit and explicitly extract the divergent part. By using this propagator,
we evaluate the part in the vacuum expectation value of the energy-momentum
tensor induced by the conical structure and the magnetic flux. In section %
\ref{sec:Gen} we consider the general case for the parameters of the conical
structure and the flux. Aiming to investigate the effects induced by this
structure, we evaluate the subtracted heat kernel and the Green function.
The corresponding part in the vacuum expectation value of the
energy-momentum tensor is presented as a sum of two terms. The first one
corresponding to the contribution coming from the bulk, having only a
three-dimensional point-like global monopole on it, and the second one is
induced by the conical structure of its two-dimensional submanifold. The
behavior of the latter in various asymptotic regions of the parameters is
discussed. The special case is discussed when the global monopole is absent.
The main results of the paper are summarized in section \ref{conc}.
Throughout the paper the system of units $\hbar =c=1$ is used.

\section{Heat kernel}

\label{sec:heat}

The six-dimensional space-time that we want to investigate the fermionic
vacuum polarization effects, is constituted by a point-like global monopole
on a three-brane and a transverse conical two-dimensional space. In the
coordinate system $x^{A}=(t,\ r,\theta ,\ \phi ,\ \rho ,\ \varphi )$ it can
be described by the line element
\begin{equation}
ds^{2}=g_{AB}dx^{A}dx^{B}=-dt^{2}+dr^{2}/\alpha ^{2}+r^{2}(d\theta ^{2}+\sin
^{2}\theta d\phi ^{2})+d\rho ^{2}+b^{2}\rho ^{2}d\varphi ^{2}\ ,  \label{com}
\end{equation}%
with radial coordinates varying as $r,\ \rho \ \geqslant 0$, planar angles $%
\phi ,\ \varphi \in \ [0,\ 2\pi ]$, polar angle $\theta \in \lbrack 0,\ \pi
] $ and $t\in (-\infty ,\ \infty )$. In (\ref{com}), the parameters $\alpha $
and $b$ codify the presence of the global monopole and string respectively.
In this space-time the scalar curvature outside the string axis is $\mathcal{%
R}=2(1-\alpha ^{2})/r^{2}$.

The flat six-dimensional Dirac matrices, $\Gamma ^{(M)}$, are $8\times 8$
matrices, which can be constructed from the four-dimensional $4\times 4$
ones \cite{B-D} as shown below \cite{Moha}:
\begin{equation}
\Gamma ^{(\mu )}=\left(
\begin{array}{cc}
0 & \gamma ^{\mu } \\
\gamma ^{\mu } & 0%
\end{array}%
\right) \ ,\ \Gamma ^{(4)}=\left(
\begin{array}{cc}
0 & i\gamma _{5} \\
i\gamma _{5} & 0%
\end{array}%
\right) \ ,\ \Gamma ^{(5)}=\left(
\begin{array}{cc}
0 & I \\
-I & 0%
\end{array}%
\right) \ ,  \label{Gamu}
\end{equation}%
where $\mu =0,1,2,3$, $\gamma _{5}=i\gamma ^{0}\gamma ^{1}\gamma ^{2}\gamma
^{3}$, and $I$ is the $4\times 4$ identity matrix. These matrices obey the
Clifford algebra $\{\Gamma ^{(M)},\ \Gamma ^{(N)}\}=-2\eta ^{(M)(N)}I_{(8)}$%
, for $M,\ N=0,\ 1,\ ...\ 5$. In this representation, the $\Gamma ^{(7)}$
matrix takes a diagonal form below
\begin{equation}
\Gamma ^{(7)}=\Gamma ^{(0)}\Gamma ^{(1)}\ ...\Gamma ^{(5)}=\left(
\begin{array}{cc}
I & 0 \\
0 & -I%
\end{array}%
\right) \ ,  \label{Ga7}
\end{equation}%
and presents two chiral eigenstates defined as $\Psi _{+}$ and $\Psi _{-}$
\cite{Moha}. So any six-dimensional fermionic wave-function, $\Psi $, can be
decomposed in terms of its chiral components as
\begin{equation}
\Psi =\left(
\begin{array}{c}
\Psi _{+} \\
\Psi _{-}%
\end{array}%
\right) \ .  \label{WavFunc}
\end{equation}

Solutions for the Dirac equation,
\begin{equation}
i\Gamma ^{(M)}\partial _{(M)}\Psi =M\Psi \ ,  \label{Deq}
\end{equation}%
with defined chirality can only be possible for $M=0$. In this way for
positive chirality, equation (\ref{Deq}) reduces to
\begin{equation}
\sigma ^{(M)}\partial _{(M)}\Psi _{+}=0\ ,  \label{DEpl}
\end{equation}%
being $\sigma ^{(M)}=(\gamma ^{\mu },\ i\gamma _{5},\ -I)$ a set of $4\times
4$ matrices; and for negative chirality it reduces to
\begin{equation}
{{\tilde{\sigma}}}^{(M)}\partial _{(M)}\Psi _{-}=0\ ,  \label{DEmin}
\end{equation}%
being now ${\tilde{\sigma}}^{(M)}=(\gamma ^{\mu },\ i\gamma _{5},\ I)$.

For a massive charged fermionic filed in a curved space-time and in the
presence of the electromagnetic field, the Dirac equation has the form
\begin{equation}
\left( i{\not{\hspace{-0.4ex}}\nabla }+e{\not{\hspace{-0.8ex}}A}-M\right)
\Psi (x)=0\ ,  \label{DEcurv}
\end{equation}%
with the covariant derivative operator defined by the relation%
\begin{equation}
{\not{\hspace{-0.4ex}}\nabla }=\Gamma ^{M}(\partial _{M}+\Pi _{M})\ ,
\label{CovDer}
\end{equation}%
where $\Pi _{M}$ is the spin connection given in terms of the $\Gamma $
matrices by
\begin{equation}
\Pi _{M}=-\frac{1}{4}\Gamma _{N}\nabla _{M}\Gamma ^{N}\ ,  \label{Pimu}
\end{equation}%
and
\begin{equation*}
{\not{\hspace{-0.8ex}}A}=\Gamma ^{M}A_{M}\ .
\end{equation*}

Now, in order to write the Dirac equation in the six-dimensional space-time
defined by (\ref{com}), we shall choose the following basis tetrad below:
\begin{equation}
e_{(M)}^{A}=\left(
\begin{array}{cccccc}
1 & 0 & 0 & 0 & 0 & 0 \\
0 & \alpha \sin \theta \cos \phi & \cos \theta \cos \phi /r & -\sin \phi
/r\sin \theta & 0 & 0 \\
0 & \alpha \sin \theta \sin \phi & \cos \theta \sin \phi /r & \cos \phi
/r\sin \theta & 0 & 0 \\
0 & \alpha \cos \theta & -\sin \theta /r & 0 & 0 & 0 \\
0 & 0 & 0 & 0 & \cos \varphi & -\sin \varphi /b\rho \\
0 & 0 & 0 & 0 & \sin \varphi & \cos \varphi /b\rho%
\end{array}%
\right) \ .  \label{e}
\end{equation}%
For this basis tetrad, the only nonzero spin connections are:
\begin{eqnarray}
\Pi _{\theta } &=&\frac{i}{2}(1-\alpha ){\vec{\Sigma}}_{(8)}\cdot {\hat{\phi}%
\,,}  \notag \\
\Pi _{\phi } &=&-\frac{i}{2}(1-\alpha )\sin \theta {\vec{\Sigma}}_{(8)}\cdot
{\hat{\theta}\,,}  \label{PiComp} \\
\Pi _{\varphi } &=&\frac{i}{2}(1-b)\Lambda _{5}\ ,  \notag
\end{eqnarray}%
where ${\hat{\theta}}$ and ${\hat{\phi}}$ are the standard unit vectors
along the angular directions on the three-brane,
\begin{equation}
{\vec{\Sigma}}_{(8)}=\left(
\begin{array}{cc}
{\vec{\Sigma}} & 0 \\
0 & {\vec{\Sigma}}%
\end{array}%
\right) \ \mathrm{with}\ \ {\vec{\Sigma}}=\left(
\begin{array}{cc}
{\vec{\sigma}}_{P} & 0 \\
0 & {\vec{\sigma}}_{P}%
\end{array}%
\right) \ ,  \label{Sig8}
\end{equation}%
with $\sigma _{P}^{k}$ being the Pauli matrices, and
\begin{equation}
\Lambda _{5}=i\ \Gamma ^{(4)}\ \Gamma ^{(5)}=\left(
\begin{array}{cc}
\gamma _{5} & 0 \\
0 & -\gamma _{5}%
\end{array}%
\right) \ .  \label{Lamb5}
\end{equation}

The fermionic propagator defined by \cite{BD}
\begin{equation}
i{\mathcal{S}}_{F}(x,x^{\prime })=\langle 0|T(\Psi (x){\bar{\Psi}}(x^{\prime
}))|0\rangle \ ,  \label{FermProp}
\end{equation}%
with ${\bar{\Psi}}=\Psi ^{\dagger }\Gamma ^{0}$, satisfies the
non-homogeneous linear differential equation,
\begin{equation}
\left( i{\not{\hspace{-0.4ex}}\nabla }+e{\not{\hspace{-0.8ex}}A}-M\right) {%
\mathcal{S}}_{F}(x,x^{\prime })=\frac{I_{(8)}}{\sqrt{-g}}\delta
^{6}(x-x^{\prime })\ ,  \label{Feyn}
\end{equation}%
where $g=\mathrm{det}(g_{AB})$. The Green function defined in (\ref{Feyn})
is a bispinor, i.e., it transforms as $\Psi $ at $x$ and as ${\bar{\Psi}}$
at $x^{\prime }$.

Let a bispinor ${\mathcal{D}}_{F}(x,x^{\prime })$ satisfy the differential
equation
\begin{eqnarray}
&&\left[ {\Box }-ieg^{MN}(D_{M}A_{N})+ie\Sigma
^{MN}F_{MN}-2ieg^{MN}A_{M}\nabla _{N}\right.  \notag \\
&&\left. -e^{2}g^{MN}A_{M}A_{N}-M^{2}-\mathcal{R}/4\right] {\mathcal{D}}%
_{F}(x,x^{\prime })=-\frac{I_{(8)}}{\sqrt{-g}}\delta ^{6}(x-x^{\prime })\ ,
\label{D}
\end{eqnarray}%
with
\begin{equation}
\Sigma ^{MN}=\frac{1}{4}[\Gamma ^{M},\Gamma ^{N}]\ ,\ D_{M}=\nabla
_{M}-ieA_{M}\ ,  \label{SigMN}
\end{equation}%
$\mathcal{R}$ being the scalar curvature and the generalized d'Alembertian
operator given by
\begin{equation}
\Box =g^{MN}\nabla _{M}\nabla _{N}=g^{MN}\left( \partial _{M}\nabla _{N}+\Pi
_{M}\nabla _{N}-\{_{MN}^{S}\}\nabla _{S}\right) \ .  \label{Dalamb}
\end{equation}%
Then the spinor Feynman propagator can be written as
\begin{equation}
{\mathcal{S}}_{F}(x,x^{\prime })=\left( i{\not{\hspace{-0.4ex}}\nabla }+e{%
\not{\hspace{-0.8ex}}A}+M\right) {\mathcal{D}}_{F}(x,x^{\prime })\ .
\label{Sf}
\end{equation}

Now we apply this formalism to the system under investigation. In order to
take into account the presence of a magnetic field along the core of the
cosmic string, we may write
\begin{equation}
A_{M}=A\partial _{M}\varphi \ .  \label{AM}
\end{equation}%
Choosing the basis tetrad (\ref{e}), the operator $\mathcal{K}$, which acts
on the bispinor in (\ref{D}), reads
\begin{eqnarray}
{\mathcal{K}} &=&-\partial _{t}^{2}+\alpha ^{2}\left( \partial _{r}^{2}+%
\frac{2}{r}\partial _{r}\right) -\frac{{\vec{L}}^{2}}{r^{2}}+\partial _{\rho
}^{2}+\frac{1}{\rho }\partial _{\rho }  \notag \\
&+&\frac{1}{b^{2}\rho ^{2}}\left( \partial _{\varphi }+\Pi _{\varphi
}-ieA_{\varphi }\right) ^{2}-\frac{(1-\alpha )}{r^{2}}\left[ 1+{\vec{\Sigma}}%
_{(8)}\cdot {\vec{L}}\right] -M^{2}\ ,  \label{KCal}
\end{eqnarray}%
where $\vec{L}$ is the ordinary angular momentum operator on the
three-brane. As we can see, the above operator is expressed as diagonal
blocks of $4\times 4$ matrices.

Because it is not possible to express the fermionic propagator for a massive
field in the manifold given by (\ref{com}) in terms of known special
functions, we shall restrict our analysis for vacuum effects associated with
massless field only. In this case we shall consider the field with positive
chirality, for which the Dirac equation can be written in terms of a $%
4\times 4$ matrix differential equation
\begin{equation}
{\not{\hspace{-0.7ex}}D}\Psi _{+}=0\ ,  \label{DEpl1}
\end{equation}%
with the operator
\begin{eqnarray}
{\not{\hspace{-0.7ex}}D} &=&\gamma ^{0}\partial _{t}+\frac{1}{r}\gamma
_{r}\left( \alpha r\partial _{r}-{\vec{\Sigma}}\cdot {\vec{L}}-1+\alpha
\right) +{\vec{\sigma}}\cdot {\hat{\rho}}\ \left( \partial _{\rho }-\frac{1-b%
}{2b\rho }\right)  \notag \\
&&+\frac{1}{b\rho }{\vec{\sigma}}\cdot {\hat{\varphi}}\ \left( \partial
_{\varphi }-ieA\ \ \right) ,  \label{DO}
\end{eqnarray}%
where $\vec{\sigma}=(\gamma ^{j},\ i\gamma _{5},\ -I)$. In (\ref{DO}) we
have $\hat{r}\cdot \vec{\sigma}=\hat{r}\cdot \vec{\gamma}=\gamma _{r}$, ${%
\vec{\sigma}}\cdot {\hat{\rho}}=i\gamma _{5}\cos \varphi -I\sin \varphi $
and ${\vec{\sigma}}\cdot {\hat{\varphi}}=-i\gamma _{5}\sin \varphi -I\cos
\varphi $, being $\hat{r}$, $\hat{\rho}$ and $\hat{\varphi}$ the unit
vectors along the corresponding coordinate directions. The case of a field
with negative chirality is considered in a similar way.

The four-component Feynman propagator obeys the differential equation
\begin{equation}
i{\not{\hspace{-0.7ex}}D}S_{F}(x,x^{\prime })=\frac{I}{\sqrt{-g}}\delta
^{(6)}(x-x^{\prime })\ ,  \label{iDe}
\end{equation}%
and can be expressed in terms of the bispinor $\mathcal{D}_{F}$ by the
relation
\begin{equation}
S_{F}(x,x^{\prime })=i{\not{\hspace{-0.7ex}}D}{\mathcal{D}}_{F}(x,x^{\prime
})\ .  \label{SFtoDF}
\end{equation}%
Now ${\mathcal{D}}_{F}(x,x^{\prime })$ is a $4\times 4$ bispinor, being the
solution of the differential equation
\begin{equation}
{\bar{\mathcal{K}}}\mathcal{D}_{F}(x,x^{\prime })=-\frac{I}{\sqrt{-g}}\delta
^{(6)}(x-x^{\prime })\ ,  \label{KbarDF}
\end{equation}%
with the operator%
\begin{eqnarray}
{\bar{\mathcal{K}}} &=&-\partial _{t}^{2}+\alpha ^{2}\left( \partial
_{r}^{2}+\frac{2}{r}\partial _{r}\right) -\frac{{\vec{L}}^{2}}{r^{2}}%
+\partial _{\rho }^{2}+\frac{1}{\rho }\partial _{\rho }  \notag \\
&+&\frac{1}{b^{2}\rho ^{2}}\left( \partial _{\varphi }+\pi _{\varphi
}-ieA_{\varphi }\right) ^{2}-\frac{(1-\alpha )}{r^{2}}\left[ 1+{\vec{\Sigma}}%
\cdot {\vec{L}}\right] \ ,  \label{K}
\end{eqnarray}%
where
\begin{equation}
\pi _{\varphi }=\frac{i}{2}(1-b)\gamma _{5}\ .  \label{piphi}
\end{equation}%
The vacuum expectation value of the energy-momentum tensor can be expressed
in terms of the Euclidean Green function, which is related with the ordinary
Feynman Green function \cite{BD} by the formula $\mathcal{D}_{E}(\tau ,\vec{r%
};\tau ^{\prime },\vec{r^{\prime }})=-i\mathcal{D}_{F}(x,x^{\prime })$,
where $t=i\tau $. In the following we shall consider the Euclidean Green
function.

In order to obtain the Euclidean Green function ${\mathcal{D}}%
_{E}(x,x^{\prime })$ in explicit form, we need to have the complete set of
normalized bispinors which obey the eigenvalue equation
\begin{equation}
{\bar{\mathcal{K}}}_{E}\Psi _{\lambda }(x)=-\lambda ^{2}\Psi _{\lambda }(x)\
\ \mathrm{with}\ \ \lambda ^{2}\geqslant 0\ ,  \label{K1}
\end{equation}%
being ${\bar{\mathcal{K}}}_{E}$ the Euclidean continuation of the
differential operator given in (\ref{K}). So, we may write
\begin{equation}
{\mathcal{D}}_{E}(x,x^{\prime })=\sum_{\lambda ^{2}}\frac{\Psi _{\lambda
}(x)\Psi _{\lambda }^{\dagger }(x^{\prime })}{\lambda ^{2}}=\int_{0}^{\infty
}\ ds\sum_{\lambda ^{2}}\Psi _{\lambda }(x)\Psi _{\lambda }^{\dagger
}(x^{\prime })\ e^{-s\lambda ^{2}}\ .  \label{D2}
\end{equation}%
The eigenfunctions of (\ref{K1}) can be specified by a set of quantum
numbers associated with operators that commute with ${\bar{\mathcal{K}}}_{E}$
and among themselves. They are: $p_{\tau }=-i\partial _{\tau }$, ${\vec{J}}%
^{2}$, $J_{z}$, ${\vec{L}}^{2}$ and ${\vec{S}}^{2}$, being ${\vec{S}}=\frac{1%
}{2}{\vec{\Sigma}}$ and ${\vec{J}}={\vec{L}}+{\vec{S}}$, and $p_{\varphi
}=-i\partial _{\varphi }$ and $\gamma _{5}$. The latter two operators are
associated with the variables on the conical two-space.

On the basis of this set of operators, the normalized eigenfunctions of the
operator ${\bar{\mathcal{K}}}_{E}$ are presented in the form
\begin{eqnarray}
\Psi _{\lambda }^{(\delta )(\sigma )}(x) &=&\frac{1}{2\pi }\sqrt{\frac{%
\alpha p\beta }{2rb}}e^{-iE\tau }e^{in\varphi }J_{\nu _{\sigma }}(pr)J_{|%
\bar{\mu }_{\delta }|/b}(\beta \rho )\left(
\begin{array}{c}
\Phi _{j,m_{j}}^{(\sigma )}(\theta ,\phi ) \\
\delta \Phi _{j,m_{j}}^{(\sigma )}(\theta ,\phi )%
\end{array}%
\right) \ ,  \label{Psi} \\
\lambda ^{2} &=&E^{2}+\alpha ^{2}p^{2}+\beta ^{2}\ ,  \label{lambda}
\end{eqnarray}%
where $J_{\mu }$ represents the Bessel function of the orders
\begin{eqnarray}
\nu _{\sigma } &=&\frac{j+1/2}{\alpha }-\frac{n_{\sigma }}{2}\ ,\ \mathrm{%
with}\ n_{\sigma }=(-1)^{\sigma }\ ,\ \sigma =0,\ 1\   \notag \\
\bar{\mu }_{\delta } &=&n+\delta \frac{1-b}{2}-N-\gamma \ ,\ \mathrm{with}\
\delta =\pm 1.  \label{mudelta}
\end{eqnarray}%
In (\ref{Psi}), $\Phi _{j,m_{j}}^{(\sigma )}$ are the spinor spherical
harmonics which are eigenfunctions of the operators $\vec{L}^{2}$ and $\vec{%
\sigma}\cdot \vec{L}$ as shown below:
\begin{eqnarray}
\vec{L}^{2}\Phi _{j,m_{j}}^{(\sigma )} &=&l(l+1)\Phi _{j,m_{j}}^{(\sigma )}\
,  \notag \\
\vec{\sigma}\cdot \vec{L}\Phi _{j,m_{j}}^{(\sigma )} &=&-(1+\kappa ^{(\sigma
)})\Phi _{j,m_{j}}^{(\sigma )}\ ,  \label{Kappa}
\end{eqnarray}%
with $\kappa ^{(0)}=-(l+1)=-(j+1/2)$ and $\kappa ^{(1)}=l=j+1/2$. \footnote{%
Explicit forms of above standard functions are given in \cite{B-D}.} In (\ref%
{mudelta}) we have expressed $eA=N+\gamma $, in terms of an integer number, $%
N$, plus a fractional one, $\gamma $. As we shall see below, only the
fractional part $\gamma $ leads to non-trivial effects. The index $\sigma $
specifies two types of eigenfunctions corresponding to $l=j-n_{\sigma }/2$
being $l$ the orbital quantum number, and $\delta =\pm 1$ the corresponding
eigenvalues of the $\gamma _{5}$ matrix. Moreover, we have $E\in (-\infty ,\
\infty )$, $n=0,\ \pm 1,\ \pm 2,\ ...$, $j=1/2,\ 3/2,\ ...$ denoting the
eigenvalue of the total angular quantum number, $m_{j}=-j,\ ...,\ j$
determines its projection and $p,\ \beta \in \lbrack 0,\ \infty )$.

According to (\ref{D2}) the heat kernel is given by the expression
\begin{equation}
{\mathcal{K}}(x,x^{\prime };s)=\int_{-\infty }^{\infty }\ dE\
\int_{0}^{\infty }\ dp\int_{0}^{\infty }\ d\beta \sum_{n,\sigma ,\delta
,j,m_{j}}\Psi _{\lambda }^{(\delta )(\sigma )}(x)\Psi _{\lambda }^{(\delta
)(\sigma )\dagger }(x^{\prime })e^{-s\lambda ^{2}}\ .  \label{KCal1}
\end{equation}%
Substituting (\ref{Psi}) into the above definition, we get:
\begin{eqnarray}
{\mathcal{K}}(x,x^{\prime };s) &=&\frac{1}{32\pi ^{3/2}b\alpha }\frac{%
e^{iN\Delta \varphi }}{\sqrt{rr^{\prime }}}\frac{e^{-{\mathcal{V}}/4s\alpha
^{2}}}{s^{5/2}}\sum_{n}\sum_{\delta }I_{|\mu _{\delta }|/b}\left( \frac{\rho
\rho ^{\prime }}{2s}\right) e^{in\Delta \varphi }  \notag \\
&&\times \sum_{\sigma =0,1}\sum_{j,m_{j}}I_{\nu _{\sigma }}\left( \frac{%
rr^{\prime }}{2\alpha ^{2}s}\right) \left(
\begin{array}{cc}
C_{j,m_{j}}^{\sigma }(\Omega ,\Omega ^{\prime }) & \delta
C_{j,m_{j}}^{\sigma }(\Omega ,\Omega ^{\prime }) \\
\delta C_{j,m_{j}}^{\sigma }(\Omega ,\Omega ^{\prime }) &
C_{j,m_{j}}^{\sigma }(\Omega ,\Omega ^{\prime })%
\end{array}%
\right) \ ,  \label{Ke}
\end{eqnarray}%
where $\Delta \varphi =\varphi -\varphi ^{\prime }$, $I_{\mu }(z)$ is the
modified Bessel function,
\begin{equation}
C_{j,m_{j}}^{\sigma }(\Omega ,\Omega ^{\prime })=\Phi _{j,m_{j}}^{(\sigma
)}(\theta ,\phi )\Phi _{j,m_{j}}^{(\sigma )\dagger }(\theta ^{\prime },\phi
^{\prime })  \label{Csig}
\end{equation}%
is a $2\times 2$ matrix and
\begin{eqnarray}
\mu _{\delta } &=&n+\delta (1-b)/2-\gamma \ ,  \notag \\
\mathcal{V} &=&\alpha ^{2}(\Delta \tau ^{2}+\rho ^{2}+\rho ^{\prime
}{}^{2})+r^{2}+r^{\prime }{}^{2}\ ,  \label{mudeltanew}
\end{eqnarray}%
with $\Delta \tau =\tau -\tau ^{\prime }$. The general expression for the
Green function is obtained by integrating (\ref{Ke}), as shown below:
\begin{equation}
{\mathcal{D}}_{E}(x,x^{\prime })=\int_{0}^{\infty }ds\ {\mathcal{K}}%
(x,x^{\prime };s).  \label{DE}
\end{equation}%
For $b=1$ and $\gamma =N=0$, the summation over the quantum number $n$ in (%
\ref{Ke}) can be done explicitly by using formula from \cite{Prud} and the
result coincides with the corresponding one given in \cite{Mello1}.

\section{Special case}

\label{sec:special}

\subsection{Green function}

\label{subsec:SpGreen}

Before to construct the Green function from the heat kernel (\ref{Ke}) for
the general case of the parameters characterizing the conical structure and
the magnetic flux, here we consider a very special case which allows us to
obtain a much simpler expression. It has been shown in \cite{Sahni} that
when the parameter $q=1/b$ is an integer number, the scalar Green function
in four-dimensional cosmic string space-time can be expressed as a sum of $q$
images of the Minkowsiki space-time function. Also, recently the image
method was used in \cite{Mello3} to provide closed expressions for the
massive scalar Green functions in a higher-dimensional cosmic string
space-time. The mathematical reason for the use of image method in these
applications is because the order of the modified Bessel functions which
appear in the heat kernel becomes an integer number. As we have seen, for
the fermionic case the order of the Bessel function depends, besides on the
integer angular quantum number $n$ also on the the factor $(1-b)/2$ from the
spin connection. However, considering a charged fermionic field in the
presence of a magnetic flux running along the string, an additional term
will be present, the factor $\gamma $. In the special case
\begin{equation}
\gamma =(1-b)/2,  \label{gamSp}
\end{equation}%
the order of the Bessel function becomes an integer number and the image
method can be used to express the fermionic Green function in a closed form
\cite{Mello4}. Here, the manifold that we are considering has the structure
of a direct product of a cosmic string by a global monopole one. Accepting
the above condition on the parameters $b$ and $\gamma $, the calculation of
the vacuum energy-momentum tensor becomes much easier to be performed as we
shall see. So, although being a very special situation, the analysis of
vacuum polarization effects in this circumstance may shed light on the
qualitative behavior of these quantities for non-integer $q$. By using the
formula \cite{Prud},
\begin{equation}
\sum_{m=-\infty }^{\infty }I_{mq}(x)e^{imq\varphi }=\frac{1}{q}%
\sum_{k=0}^{q-1}e^{x\cos (\varphi +2\pi k/q)}\ ,  \label{SumBes}
\end{equation}%
after some intermediate steps, (\ref{Ke}) can be written as:
\begin{eqnarray}
{\mathcal{K}}(x,x^{\prime };s) &=&\frac{1}{32\pi ^{3/2}\alpha }\frac{%
e^{iN\Delta \varphi }}{\sqrt{rr^{\prime }}}\sum_{k=0}^{q-1}\frac{e^{-{%
\mathcal{V}}_{k}/4s\alpha ^{2}}}{s^{5/2}}\sum_{\sigma
=0,1}\sum_{j,m_{j}}I_{\nu _{\sigma }}\left( \frac{rr^{\prime }}{2\alpha ^{2}s%
}\right)  \notag \\
&&\times \left(
\begin{array}{cc}
(1+e^{i\beta _{k}})C_{j,m_{j}}^{\sigma }(\Omega ,\Omega ^{\prime }) &
(1-e^{i\beta _{k}})C_{j,m_{j}}^{\sigma }(\Omega ,\Omega ^{\prime }) \\
(1-e^{i\beta _{k}})C_{j,m_{j}}^{\sigma }(\Omega ,\Omega ^{\prime }) &
(1+e^{i\beta _{k}})C_{j,m_{j}}^{\sigma }(\Omega ,\Omega ^{\prime })%
\end{array}%
\right) \ ,  \label{Ks}
\end{eqnarray}%
with notations
\begin{eqnarray}
{\mathcal{V}}_{k} &=&{\mathcal{V}}-2\alpha ^{2}\rho \rho ^{\prime }\cos
\left( \Delta \varphi /q+2k\pi /q\right) ,  \notag \\
\beta _{k} &=&(1-1/q)\Delta \varphi -2k\pi /q\ .  \label{betk}
\end{eqnarray}

The corresponding Green function is obtained by substituting (\ref{Ks}) into
(\ref{DE}). After the integration over the variable $s$ with the help of
formula from \cite{Prud}, the result is:
\begin{eqnarray}
{\mathcal{D}}_{E}(x,x^{\prime }) &=&-\frac{\alpha ^{2}}{8\pi ^{2}}\frac{%
e^{iN\Delta \varphi }}{(r^{\prime }r)^{2}}\sum_{k=0}^{q-1}\sum_{\sigma
=0,1}\sum_{j,m_{j}}\frac{Q_{\nu _{\sigma }-1/2}^{1}(\cosh u_{k})}{\sinh u_{k}%
}  \notag \\
&&\times \left(
\begin{array}{cc}
(1+e^{i\beta _{k}})C_{j,m_{j}}^{\sigma }(\Omega ,\Omega ^{\prime }) &
(1-e^{i\beta _{k}})C_{j,m_{j}}^{\sigma }(\Omega ,\Omega ^{\prime }) \\
(1-e^{i\beta _{k}})C_{j,m_{j}}^{\sigma }(\Omega ,\Omega ^{\prime }) &
(1+e^{i\beta _{k}})C_{j,m_{j}}^{\sigma }(\Omega ,\Omega ^{\prime })%
\end{array}%
\right) \ ,  \label{Desp}
\end{eqnarray}%
with
\begin{equation*}
\cosh u_{k}=\frac{{\mathcal{V}}_{k}}{2r^{\prime }r}=\frac{\alpha ^{2}(\Delta
\tau ^{2}+\rho ^{2}+\rho ^{\prime 2}-2\rho \rho ^{\prime }\cos (\Delta
\varphi /q+2\pi k/q))+r^{\prime 2}+r^{2}}{2r^{\prime }r}\ .
\end{equation*}%
In (\ref{Desp}), $Q_{\nu }^{l}(z)$ represents the associated Legendre
function of the second kind. From the above expression we can see that, the $%
k=0$ component presents divergence at the coincidence limit; however, as to
the others, they remain finite. The reason is because $\cosh u_{k}$ will be
always greater than unity for these components, consequently the Legendre
function assumes finite value.

\subsection{Vacuum average of the energy-momentum tensor}

\label{subsec:SpEMT}

The evaluation of the vacuum expectation value (VEV) of the energy-momentum
tensor associated with massless fermionic field in the six-dimensional
space, in a braneworld scenario containing a three-dimensional global
monopole, has been developed in \cite{Mello1}. Here we are mainly interested
in the investigation of quantum effects induced by the presence of the
cosmic string in the transverse two-dimensional space and by the magnetic
flux along the core of the string.

Using the point-splitting procedure \cite{BD}, the VEV of the
energy-momentum tensor associated with charged fermionic fields can be
obtained by using the Feynmann propagator as shown below:
\begin{equation}
\langle T_{AB}(x)\rangle =\frac{1}{4}\lim_{x^{\prime }\rightarrow x}\mathrm{%
Tr}\left[ {\tilde{\sigma}}_{A}(D_{B}-{\bar{D}}_{B^{\prime }})+{\tilde{\sigma}%
}_{B}(D_{A}-{\bar{D}}_{A^{\prime }})\right] S_{F}(x,x^{\prime })\ .
\label{EM}
\end{equation}%
Here we shall use the four-component bispinor, where $D_{B}=\nabla
_{B}-iA_{B}$, the bar denotes complex conjugate, and ${\tilde{\sigma}}%
_{A}=e_{A}^{(M)}\ {\tilde{\sigma}}_{(M)}$, with ${\tilde{\sigma}}%
^{(M)}=(\gamma ^{\mu },\ i\gamma _{5},\ I)$. Moreover,
\begin{equation}
S_{F}(x,x^{\prime })=i{\not{\hspace{-0.7ex}}D}{\mathcal{D}}_{F}(x,x^{\prime
})\ ,  \label{SFsp}
\end{equation}%
being the operator ${\not{\hspace{-0.8ex}}D}$ given by (\ref{DO}).

For the special case considered in this section, we shall use for the
bispinor the result given in (\ref{Desp}). As the first step in this
direction, let us calculate the zero-zero component of (\ref{EM}). Because ${%
\tilde{\sigma}}_{0}=-\gamma ^{0}$, $D_{0}=\partial _{0}$ and the dependence
of the Feynman propagator on the time variable is by $t-t^{\prime }$, we can
write:
\begin{equation}
\langle T_{00}(x)\rangle =\lim_{x^{\prime }\rightarrow x}\mathrm{Tr}\left[
\gamma _{0}\partial _{0}{\mathcal{S}}_{F}(x^{\prime },x)\right]
=-\lim_{x^{\prime }\rightarrow x}\mathrm{Tr}\left[ \gamma _{0}\partial _{0}{%
\not{\hspace{-0.7ex}}D}{\mathcal{D}}_{E}(x^{\prime },x)\right] \ ,
\label{T00s1}
\end{equation}%
where we have used the Euclidean version for the Green function.

Before to embark in the above calculation it is useful to analyze,
separately, the contribution coming from the $k=0$ component of the
Euclidean Green function. As the first point we can see that taking $\Omega
=\Omega ^{\prime }$ in $C_{j,m_{j}}^{\sigma }$, the summation over the
quantum number $m_{j}$ in (\ref{Desp}) provides:
\begin{equation}
\sum_{m_{j}}C_{j,m_{j}}^{\sigma }(\Omega ,\Omega )=\frac{2j+1}{8\pi }%
I_{(2)}\ .  \label{Summj}
\end{equation}%
On the other hand, taking $\varphi ^{\prime }=\varphi $, the $k=0$ component
of the Green function becomes proportional to $I_{(4)}$. The same is true
when the operator ${\vec{\Sigma}}\cdot {\vec{L}}$ of (\ref{DO}) acts on $%
C_{j,m_{j}}^{\sigma }$, because of (\ref{Kappa}). In this way, almost all
derivative terms provide a vanishing trace. The exception is for the term
proportional to $\partial _{\varphi }$; however, after taking this azimuthal
angle derivative on (\ref{Desp}) and the coincidence limit $\varphi ^{\prime
}=\varphi $, the matrix left is proportional to $i(1-b)(I_{(4)}-\gamma _{5})$%
, consequently its products with $\gamma _{5}\gamma _{0}$ and $\gamma _{0}$
have a vanishing trace. As a conclusion we can affirm that the only term
that produces a non-vanishing result in the calculation of the vacuum
average is linear in the time derivative in (\ref{DO}), consequently for the
$k=0$ component we have:
\begin{equation}
\langle T_{00}(x)\rangle ^{(k=0)}=-\lim_{x^{\prime }\rightarrow x}\mathrm{Tr}%
\ \partial _{\tau }^{2}\ {\mathcal{D}}_{E}^{(k=0)}(x,x^{\prime })\ .
\label{T00k0}
\end{equation}%
Taking in ${\mathcal{D}}_{E}^{(k=0)}(x,x^{\prime })$ the coincidence limit
for the azimuthal angle in the extra conical two-dimensional space, this
component coincides with the corresponding bispinor in the absence of string
and magnetic flux. As a consequence the analysis of the above vacuum average
is similar to that one developed in \cite{Mello1}. The new analysis will
involve the $k\geqslant 1$ components of (\ref{Desp}). As we have already
mentioned, the main objective of this paper is to evaluate the quantum
effects induced by the string and magnetic flux on the VEV of the fermionic
energy-momentum tensor. So, our next step is to calculate the contributions
from the string and magnetic flux to $\langle T_{00}(x)\rangle $. We will
denote this part as $\langle T_{00}(x)\rangle _{\mathrm{s}}$.

Because the $k\geqslant 1$ components of (\ref{Desp}) are finite at the
coincidence limit for $\rho \neq 0$, the calculation of their contribution
to the VEV of the zero-zero component of the energy-momentum tensor does not
require any renormalization procedure. Moreover, this contribution is only
due to the term of the operator (\ref{DO}) with time derivative . The reason
resides in the fact that only second time derivatives of $1/\sinh u_{k}$ and
$Q_{\nu _{\sigma }-1/2}^{1}(\cosh u_{k})$ in (\ref{Desp}), present
non-vanishing result in the coincidence limit. So for these component of the
bispinor we also can write:
\begin{equation}
\langle T_{00}(x)\rangle _{\mathrm{s}}=-\lim_{x^{\prime }\rightarrow x}%
\mathrm{Tr}\left[ \gamma _{0}\partial _{0}{\not{\hspace{-0.7ex}}D}{\mathcal{D%
}}_{E}^{(k\geqslant 1)}(x,x^{\prime })\right] =-\lim_{x^{\prime }\rightarrow
x}\mathrm{Tr}\ \partial _{\tau }^{2}\ {\mathcal{D}}_{E}^{(k\geqslant
1)}(x,x^{\prime })\ .  \label{T00s}
\end{equation}%
Now, after some intermediate steps and introducing the rescaled radial
coordinate $\tilde{r}=r/\alpha $, we arrive at the expression:
\begin{equation}
\langle T_{0}^{0}\rangle _{\mathrm{s}}=-\frac{\alpha ^{-2}}{16\pi ^{3}\tilde{%
r}^{6}y}\sum_{k=1}^{q-1}\frac{\cot ^{2}(\pi k/q)}{1+y\sin ^{2}(\pi k/q)}%
\sum_{l=1}^{\infty }l\left[ Q_{l/\alpha -1}^{2}(z_{k})+Q_{l/\alpha
}^{2}(z_{k})\right] \,,  \label{tt}
\end{equation}%
where
\begin{equation}
y=\left( \rho /\tilde{r}\right) ^{2},\ \ z_{k}=1+2y\sin ^{2}(\pi k/q)\ .
\label{y}
\end{equation}%
By making use of the recurrence relation for the associated Legendre
function of the second kind (see \cite{Abra64}), it can be seen that%
\begin{equation}
Q_{\nu -1}^{2}(z)+Q_{\nu }^{2}(z)=-2\sqrt{\frac{z+1}{z-1}}\left[ (1+\nu
)Q_{\nu -1}^{1}(z)+(1-\nu )Q_{\nu }^{1}(z)\right] ,  \label{Qrel}
\end{equation}%
and, hence, the summand in (\ref{tt}) can also be written in terms of the
associated Legendre function of the order 1.

When the global monopole is absent one has $\alpha =1$, and by using the
formula
\begin{equation}
\sum_{l=0}^{\infty }(2l+1)Q_{l}^{2}(x)=2\frac{x+1}{(x-1)^{2}},  \label{Qsum}
\end{equation}%
from (\ref{tt}) we find

\begin{equation}
\langle T_{0}^{0}\rangle _{\mathrm{s}}=-\frac{1}{16\pi ^{3}\rho ^{6}}%
\sum_{k=1}^{q-1}\frac{\cos ^{2}(\pi k/q)}{\sin ^{6}(\pi k/q)}=-\frac{q^{2}-1%
}{7560\pi ^{3}\rho ^{6}}(q^{2}-4)(q^{2}+5).  \label{TooStr}
\end{equation}%
In section \ref{sec:Gen} we shall see that these result holds for
non-integer values of $q$ as well. By taking into account that the same
result will be obtained for the negative chirality, in the case of the
combination of both chiralities formula (\ref{TooStr}) coincides with the
result given in \cite{Mello4}.

\begin{figure}[tbph]
\begin{center}
\begin{tabular}{cc}
\epsfig{figure=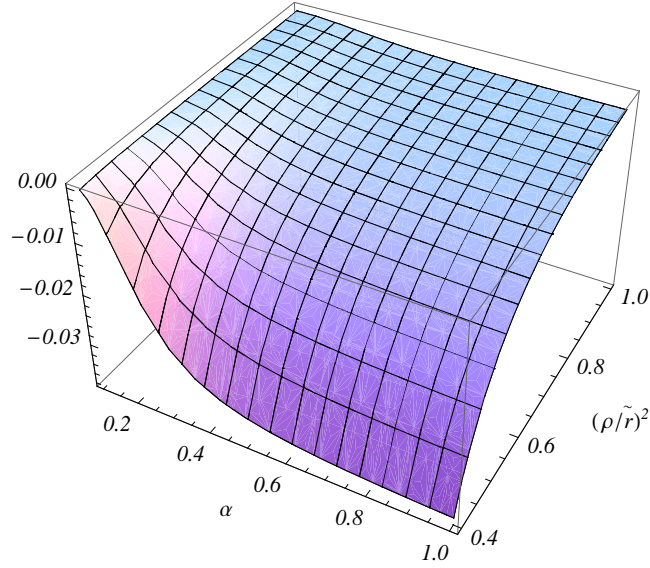, width=8.5cm, height=7.5cm} &
\end{tabular}%
\end{center}
\caption{The energy density induced by the string,
$\tilde{r}^{6}\langle {T}_{0}^{0}\rangle _{\mathrm{s}}$, as a
function of $\protect\alpha $ and $y=(\protect\rho
/\tilde{r})^{2}$ for $q=3$, $\protect\gamma =1/3$.} \label{Fig1}
\end{figure}

For $q=2$, the expression in (\ref{tt}) vanishes, and there is no
contribution to the VEV of the energy density induced by the cosmic string
and magnetic flux. However, for $q=3$ there appears a non-vanishing
contribution given below:
\begin{equation}
\langle T_{0}^{0}\rangle _{\mathrm{s}}=-\frac{\alpha ^{-2}}{24\pi ^{3}\tilde{%
r}^{6}y}\frac{1}{1+3y/4}\sum_{l=1}^{\infty }l\left[ Q_{l/\alpha
-1}^{2}\left( 1+3y/2\right) +Q_{l/\alpha }^{2}\left( 1+3y/2\right) \right] \
.  \label{T00Spq3}
\end{equation}%
In figure \ref{Fig1}, we have plotted the behavior of $\tilde{r}^{6}\langle
T_{0}^{0}\rangle _{\mathrm{s}}$, evaluated by formula (\ref{T00Spq3}), as a
function of the parameters $\alpha $ and $y$. As it is seen from this
figure, the part in the VEV of the energy density tends to zero for small
values of the parameter $\alpha $. The behavior of the VEV in various
asymptotic regions of the parameters will be described below for the general
case when there is no relation between the planar angle deficit and the
fractional part of the magnetic flux.

\begin{figure}[tbph]
\begin{center}
\begin{tabular}{cc}
\epsfig{figure=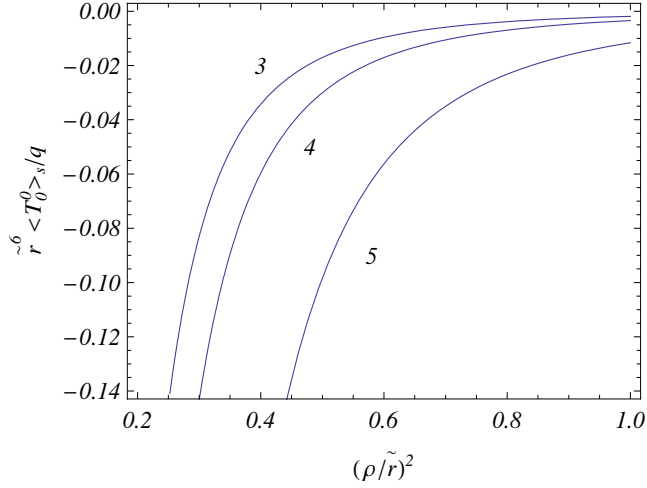, width=8.5cm, height=6.5cm} &
\end{tabular}%
\end{center}
\caption{The dependence of the energy density induced by the string, $\tilde{%
r}^{6}\langle T_{0}^{0}\rangle _{\mathrm{s}}/q$, as a function of $y=(%
\protect\rho /\tilde{r})^{2}$ for $\protect\alpha =0.5$. The
numbers near the curves correspond to the values of the
parameter~$q$.} \label{Fig2}
\end{figure}

In figure \ref{Fig2}, we have presented the dependence of the part
in the energy density induced by the string, on $y=(\rho
/\tilde{r})^{2}$ for different values of the parameter $q$ (the
numbers near the curves). The
corresponding values of the parameter $\gamma $ are found from condition (%
\ref{gamSp}). The graphs are plotted for the value of the global monopole
parameter $\alpha =0.5$. Note that in the cases $q=1,2$ the corresponding
energy density vanishes.

The VEV of the radial component of the energy-momentum tensor will be given
by
\begin{equation}
\langle T_{rr}(x)\rangle =\frac{1}{2\alpha }\lim_{x^{\prime }\rightarrow x}%
\mathrm{Tr}\left[ {\hat{r}}\cdot {\vec{\gamma}}\left( \partial _{r}-\partial
_{r^{\prime }}\right) \right] S_{F}(x,x^{\prime })\ ,  \label{Trr0}
\end{equation}%
where we have used $D_{r}=\partial _{r}$ and ${\tilde{\sigma}}_{r}={\hat{r}}%
\cdot {\vec{\gamma}}/\alpha $. Expressing the Feynman propagator in terms of
the Euclidean Green function, we can write:
\begin{equation}
\langle T_{rr}(x)\rangle =-\frac{1}{2\alpha }\lim_{x^{\prime }\rightarrow x}%
\mathrm{Tr}\left[ {\hat{r}}\cdot {\vec{\gamma}}\left( \partial _{r}-\partial
_{r^{\prime }}\right) {\not{\hspace{-0.7ex}}D}{\mathcal{D}}_{E}(x^{\prime
},x)\right] \ .  \label{Trr11}
\end{equation}%
At this point we can make a similar analysis as we did in the calculation of
the VEV of the energy density. For the $k=0$ component of (\ref{Desp}), the
only non-vanishing contributions coming from the product of $\gamma ^{r}$ by
(\ref{DO}), are given by the second, third and fourth terms of the latter.
So, we may take for this component $\varphi ^{\prime }=\varphi $. As a
consequence the corresponding vacuum average again coincides with the one in
the absence of string and magnetic flux. So, let us now proceed the
calculation of the new contributions for the above VEV due to the terms with
$k\geqslant 1$ in (\ref{Desp}):
\begin{equation}
\langle T_{rr}(x)\rangle _{\mathrm{s}}=-\frac{1}{2\alpha }\lim_{x^{\prime
}\rightarrow x}\mathrm{Tr}\left[ {\hat{r}}\cdot {\vec{\gamma}}\left(
\partial _{r}-\partial _{r^{\prime }}\right) {\not{\hspace{-0.7ex}}D}{%
\mathcal{D}}_{E}^{(k\geqslant 1)}(x,x^{\prime })\right] \ .  \label{Trr12}
\end{equation}

Because of some simplifications that appear in this calculation, it is
possible to observe that the only non-vanishing contribution comes from the
second term on the right-hand side of (\ref{DO}) and one finds
\begin{equation}
\langle T_{rr}(x)\rangle _{\mathrm{s}}=\frac{1}{2}\lim_{x^{\prime
}\rightarrow x}\mathrm{Tr}\left[ \partial _{r}(\partial _{r}-\partial
_{r^{\prime }})-\frac{\alpha +\kappa ^{(\sigma )}}{\alpha r^{2}}\right] {%
\mathcal{D}}_{E}^{(k\geqslant 1)}(x,x^{\prime })\ .  \label{trrn}
\end{equation}%
Now substituting (\ref{Desp}) into (\ref{trrn}), and using relation (\ref%
{Qrel}), we can see that
\begin{equation}
\langle T_{r}^{r}\rangle _{\mathrm{s}}=\langle T_{0}^{0}\rangle _{\mathrm{s}%
}.  \label{TrrT00}
\end{equation}

The calculation of the component $\langle T_{\rho \rho }\rangle $ is
formally given by
\begin{equation}
\langle T_{\rho \rho }\rangle =\frac{1}{2}\lim_{x^{\prime }\rightarrow x}%
\mathrm{Tr}\left[ {\tilde{\sigma}}_{\rho }(\partial _{\rho }-\partial _{\rho
^{\prime }})\right] S_{F}(x,x^{\prime })\ ,  \label{Troros}
\end{equation}%
where we have used $D_{\rho }=\partial _{\rho }$. Expressing the Feynman
propagator in terms of the Green function, and analyzing the $k=0$ component
of (\ref{Desp}), we conclude, after a careful analysis, that this component
provides the same vacuum average as in the absence of string and magnetic
flux. The only non-vanishing contributions to the parts in the VEV due to
the string and magnetic flux come from the last two terms on the right-hand
side of (\ref{DO}):
\begin{equation}
\langle T_{\rho }^{\rho }(x)\rangle _{\mathrm{s}}=\frac{1}{2}\lim_{x^{\prime
}\rightarrow x}\mathrm{Tr}\left[ \partial _{\rho }(\partial _{\rho
}-\partial _{\rho ^{\prime }})+\frac{q-1}{2\rho ^{2}}-\frac{q\gamma _{5}}{%
\rho ^{2}}(i\partial _{\varphi }+{\bar{N}}+\gamma )\right] {\mathcal{D}}%
_{E}^{(k\geqslant 1)}(x,x^{\prime })\ ,  \label{Troros1}
\end{equation}%
where $\gamma =(q-1)/2q$. Substituting the expression for the Green
function, and after a long calculation, we obtain for VEV of $T_{\rho
}^{\rho }$ the same expression as obtained for $T_{0}^{0}$, given in (\ref%
{tt}):%
\begin{equation}
\langle T_{\rho }^{\rho }\rangle _{\mathrm{s}}=\langle T_{0}^{0}\rangle _{%
\mathrm{s}}.  \label{TroroT00}
\end{equation}

The other components of the VEV for the energy-momentum tensor are found
from the covariant continuity equation $\nabla _{A}\langle T_{B}^{A}\rangle
=0$. For the geometry under consideration this equation reduces to the
following two relations
\begin{eqnarray}
\langle T_{\theta }^{\theta }\rangle &=&\langle T_{\phi }^{\phi }\rangle
=\left( 1+\frac{r}{2}\partial _{r}\right) \langle T_{r}^{r}\rangle \ ,
\notag \\
\langle T_{\varphi }^{\varphi }\rangle &=&\partial _{\rho }(\rho \langle
T_{\rho }^{\rho }\rangle )\ .  \label{ContEq}
\end{eqnarray}%
By using formulae (\ref{TrrT00}), (\ref{TroroT00}), (\ref{ContEq}), it can
be checked that the part in the VEV of the energy-momentum tensor induced by
the string and magnetic flux is traceless, $\langle T_{A}^{A}\rangle _{%
\mathrm{s}}=0$, and the trace anomaly is contained in the pure global
monopole part only.

Having obtained the complete expressions for the VEVs of all components of
the fermionic energy-momentum tensor induced by the cosmic string and
magnetic flux, we let for the next section the analysis of the general case,
where there is no more relation between the magnetic flux and the parameter
associated with the cosmic string.

\section{ General case}

\label{sec:Gen}

\subsection{Green function}

\label{subsec:GenGreen}

In order to evaluate the vacuum expectation value of the energy-momentum
tensor induced by the magnetic flux and cosmic string for general values of
the parameters $b$ and $\gamma $, we shall subtract from the heat kernel (%
\ref{Ke}) the corresponding heat kernel in the absence of cosmic string and
taking $\gamma =0$. \footnote{%
It is possible to see that the expression obtained from (\ref{Ke}) by taking
$b=1$ and $\gamma =0$ coincides, up to the factor $e^{iN \Delta \varphi }$
with the expression for the heat kernel given in \cite{Mello1}} Denoting by $%
{\mathcal{K}}_{\mathrm{sub}}(x,x^{\prime };s)$ the subtracted heat kernel
and introducing rescaled coordinates $\tilde{\varphi}=b\varphi $ and $\tilde{%
r}=r/\alpha $, we have
\begin{eqnarray}
{\mathcal{K}}_{\mathrm{sub}}(x,x^{\prime };s) &=&\frac{\alpha ^{-2}}{32\pi
^{3/2}\sqrt{\tilde{r}\tilde{r}^{\prime }}}\frac{e^{-\tilde{{\mathcal{V}}}/4s}%
}{s^{5/2}}\sum_{\sigma =0,1}\sum_{j,m_{j}}I_{\nu _{\sigma }}\left( \frac{%
\tilde{r}\tilde{r}^{\prime }}{2s}\right)  \notag \\
&&\times \sum_{n}\sum_{\delta }\left[ qe^{iNq\Delta \tilde{\varphi}}I_{q|\mu
_{\delta }|}\left( \frac{\rho \rho ^{\prime }}{2s}\right) e^{iqn{\Delta }%
\tilde{\varphi}}-I_{n}\left( \frac{\rho \rho ^{\prime }}{2s}\right) e^{in{%
\Delta }\tilde{\varphi}}\right]  \notag \\
&&\times \left(
\begin{array}{cc}
C_{j,m_{j}}^{\sigma }(\Omega ,\Omega ^{\prime }) & \delta
C_{j,m_{j}}^{\sigma }(\Omega ,\Omega ^{\prime }) \\
\delta C_{j,m_{j}}^{\sigma }(\Omega ,\Omega ^{\prime }) &
C_{j,m_{j}}^{\sigma }(\Omega ,\Omega ^{\prime })%
\end{array}%
\right) \ ,  \label{Ksub}
\end{eqnarray}%
where ${\Delta }\tilde{\varphi}=\tilde{\varphi}-\tilde{\varphi}^{\prime }$
and
\begin{equation}
\tilde{\mathcal{V}}=\Delta \tau ^{2}+\rho ^{2}+\rho ^{\prime }{}^{2}+\tilde{r%
}^{2}+\tilde{r}^{\prime }{}^{2}\ .  \label{Vtilde}
\end{equation}%
Though the contribution of the separate terms in square brackets of (\ref%
{Ksub}) to the Green function is divergent in the coincidence limit, the
part of the Green function corresponding to the subtracted heat kernel (\ref%
{Ksub}) is finite in the coincidence limit for points away from the string
core. This follows also from general arguments. For points outside the
string core the electromagnetic field is zero and the local geometry is the
same as that in the case when the global monopole is present only. Hence,
the divergences in the Green function in the coincidence limit are the same
in these two situations.

To provide a more convenient expression for the subtracted heat kernel, for
the summation over $n$ we apply the Abel-Plana formula in the form
\begin{equation}
\sum_{n=n_{\pm }}^{\infty }F(n\pm \beta )=\int_{0}^{\infty
}du\,F(u)+i\int_{0}^{\infty }du\sum_{\lambda =\pm 1}\frac{\lambda F(i\lambda
u)}{e^{2\pi (u\pm i\lambda \beta )}-1},  \label{AbelPl}
\end{equation}%
where $n_{+}=0$, $n_{-}=1$, and $0<\beta <1$. This generalization of the
Abel-Plana formula with the upper sign is given in \cite{Inui03} and the
formula with the lower sign is easily obtained from the upper sign case by
redefining $n\rightarrow n+1$. Note that formula (\ref{AbelPl}) is also
obtained from the generalized Abel-Plana formula (see \cite{Saha07}).

Defining the sum%
\begin{equation}
S_{q,\gamma }^{(\delta )}(z,{\Delta }\tilde{\varphi})=e^{iNq\Delta \tilde{%
\varphi}}\sum_{n}qI_{q|\mu _{\delta }|}(z)e^{iqn{\Delta }\tilde{\varphi}},
\label{SqgamDef}
\end{equation}%
and using formula (\ref{AbelPl}), after some intermediate steps we find%
\begin{eqnarray}
S_{q,\gamma }^{(\delta )}(z,{\Delta }\tilde{\varphi}) &=&2e^{i(N-\beta
_{\delta })q{\Delta }\tilde{\varphi}}\bigg[\int_{0}^{\infty }duI_{u}(z)\cos
(u{\Delta }\tilde{\varphi})  \notag \\
&&+\frac{1}{\pi }\int_{0}^{\infty }du\sinh (u\pi )K_{iu}(z)\sum_{\lambda
=\pm 1}\frac{e^{-\mathrm{sgn}(\beta _{\delta })\lambda u{\Delta }\tilde{%
\varphi}}}{e^{2\pi (u/q+i\lambda |\beta _{\delta }|)}-1}\bigg],
\label{Sqgam}
\end{eqnarray}%
where $K_{\nu }(z)$ is the MacDonald function and we have defined%
\begin{equation}
\beta _{\delta }=\delta (q-1)/2q-\gamma .  \label{sigmadel}
\end{equation}%
To the sum over $n$ with the second term in the square brackets of (\ref%
{Ksub}) we apply the Abel-Plana formula in the standard form. This gives
\begin{eqnarray}
S(z,{\Delta }\tilde{\varphi}) &=&\sum_{n}I_{n}\left( z\right) e^{in{\Delta }%
\tilde{\varphi}}=2\int_{0}^{\infty }duI_{u}(z)\cos (u{\Delta }\tilde{\varphi}%
)  \notag \\
&&+\frac{4}{\pi }\int du\sinh (\pi u)K_{iu}(z)\frac{\cosh (u{\Delta }\tilde{%
\varphi})}{e^{2\pi u}-1}\ .  \label{S}
\end{eqnarray}

The subtracted Green function, ${\mathcal{D}}_{\mathrm{sub}}(x,x^{\prime })$%
, that we shall use to calculate the vacuum polarization effects induced by
the cosmic string and magnetic flux, is obtained from formulae (\ref{DE}) and (%
\ref{Ksub}):
\begin{eqnarray}
{\mathcal{D}}_{\mathrm{sub}}(x,x^{\prime }) &=&\frac{\sqrt{2}\alpha ^{-2}}{%
16\pi ^{3/2}\sqrt{\tilde{r}\tilde{r}^{\prime }}}\int_{0}^{\infty }du\
u^{1/2}e^{-u\tilde{{\mathcal{V}}}/2}\bigg[\sum_{\delta }S_{q,\gamma
}^{(\delta )}(u\rho \rho ^{\prime },{\Delta }\tilde{\varphi})-S(u\rho \rho
^{\prime },{\Delta }\tilde{\varphi})\bigg]  \notag \\
&&\times \sum_{\sigma =0,1}\sum_{j,m_{j}}I_{\nu _{\sigma }}\left( u\tilde{r}%
\tilde{r}^{\prime }\right) \left(
\begin{array}{cc}
C_{j,m_{j}}^{\sigma }(\Omega ,\Omega ^{\prime }) & \delta
C_{j,m_{j}}^{\sigma }(\Omega ,\Omega ^{\prime }) \\
\delta C_{j,m_{j}}^{\sigma }(\Omega ,\Omega ^{\prime }) &
C_{j,m_{j}}^{\sigma }(\Omega ,\Omega ^{\prime })%
\end{array}%
\right) ,\   \label{DsubGen}
\end{eqnarray}%
where $\tilde{{\mathcal{V}}}$ is defined by relation (\ref{Vtilde}). The
divergent contributions from the separate terms in the square brackets of (%
\ref{DsubGen}) come from the first integrals on the right-hand sides of
formulae (\ref{Sqgam}) and (\ref{S}). Now we see that after the application
of the Abel-Plana formula these contributions are explicitly cancelled out
in the subtracted Green function for ${\Delta }\tilde{\varphi}=0$.

\subsection{Energy-momentum tensor}

\label{subsec:GenEMT}

Having the subtracted Green function we can evaluate the VEV of the
energy-momentum tensor on the base of formula (\ref{EM}). As in the special
case discussed before, let us consider the part in the vacuum average of $%
T_{0}^{0}$ induced by the string and flux:
\begin{equation}
\langle T_{0}^{0}(x)\rangle _{\mathrm{s}}=\lim_{x^{\prime }\rightarrow x}%
\mathrm{Tr}\left[ \gamma _{0}\partial _{0}{\not{\hspace{-0.7ex}}D}{\mathcal{D%
}}_{\mathrm{sub}}(x,x^{\prime })\right] \ .  \label{T00Gen}
\end{equation}%
Being ${\mathcal{D}}_{\mathrm{sub}}(x,x^{\prime })$ finite at the
coincidence limit, we can interchange the differential operator and the
integral over $u$. Moreover, terms linear in time derivative acting on the
exponential factor of the subtracted heat kernel, $e^{-\tilde{{\mathcal{V}}}%
/4s}$, produce a term linear in $\Delta \tau $, which goes to zero at the
coincidence limit. The only term that survives is the one obtained by the
second time derivative. So on the basis of these information, we may write:
\begin{equation}
\langle T_{0}^{0}(x)\rangle _{\mathrm{s}}=\lim_{x^{\prime }\rightarrow x}%
\mathrm{Tr}\partial _{\tau }^{2}{\mathcal{D}}_{\mathrm{sub}}(x,x^{\prime
})=\lim_{x^{\prime }\rightarrow x}\mathrm{Tr}\int_{0}^{\infty }ds\ \partial
_{\tau }^{2}{\mathcal{K}}_{\mathrm{sub}}(x,x^{\prime };s)\ .  \label{T00Gen1}
\end{equation}%
Developing all the calculations needed and defining a new integration
variable, $v=u\tilde{r}^{2}$, we obtain:
\begin{eqnarray}
\langle T_{0}^{0}\rangle _{\mathrm{s}} &=&-\frac{\sqrt{2}\alpha ^{-2}}{16\pi
^{5/2}\tilde{r}^{6}}\int_{0}^{\infty }dv\ v^{3/2}e^{-(1+y)v}\sum_{\sigma
=0,1}\sum_{j}(j+1/2)  \notag \\
&&\times I_{\nu _{\sigma }}\left( v\right) \sum_{n}\bigg[\sum_{\delta =\pm
1}qI_{q|\mu _{\delta }|}(yv)-2I_{n}\left( yv\right) \bigg]\,.
\label{T00sGen0}
\end{eqnarray}%
with $y$ defined by (\ref{y}).

An equivalent form for $\langle T_{0}^{0}\rangle _{\mathrm{s}}$ is obtained
by using formulae (\ref{Sqgam}) and (\ref{S}) with ${\Delta }\tilde{\varphi}%
=0$:
\begin{equation}
\langle T_{0}^{0}\rangle _{\mathrm{s}}=-\frac{\sqrt{2}}{16\pi ^{5/2}\alpha
^{2}\tilde{r}^{6}}\int_{0}^{\infty }dv\ v^{3/2}e^{-(1+y)v}\sum_{l=1}^{\infty
}l\left[ I_{l/\alpha -1/2}(v)+I_{l/\alpha +1/2}(v)\right] F_{q,\gamma }(yv),
\label{T00sGen}
\end{equation}
In this formula we have introduced the notation%
\begin{equation}
F_{q,\gamma }(z)=\sum_{\delta =\pm 1}S_{q,\gamma }^{(\delta )}\left(
z,0\right) -2S\left( z,0\right) =\frac{2}{\pi }\int_{0}^{\infty }du\sinh
(u\pi )K_{iu}(z)G_{q,\gamma }(u).  \label{Fz}
\end{equation}%
where%
\begin{equation}
G_{q,\gamma }(u)=\sum_{\delta =\pm 1}\sum_{\lambda =\pm 1}\frac{1}{e^{2\pi
(u/q+i\lambda |\beta _{\delta }|)}-1}-\frac{4}{e^{2\pi u}-1},  \label{Gqgam}
\end{equation}%
and $\beta _{\delta }$ is defined by (\ref{sigmadel}). Note that,
as it follows from (\ref{Gqgam}), the VEV of the energy-momentum
tensor does not depend on the sign of the fractional part of the
magnetic flux.

We could have obtained the second relation in (\ref{Fz}) by applying to the
series over $n$ in the definitions of $S_{q,\gamma }^{(\delta )}\left(
z,0\right) $ and $S\left( z,0\right) $ the summation formula
\begin{equation}
\sum_{n=-\infty }^{\infty }F(|n+\beta |)=2\int_{0}^{\infty
}du\,F(u)+i\int_{0}^{\infty }du\sum_{\lambda =\pm 1}\frac{F(iu)-F(-iu)}{%
e^{2\pi (u+i\lambda \beta )}-1},  \label{AbelPl2}
\end{equation}%
which directly follows from (\ref{AbelPl}). Note that for this formula the
restriction $0<\beta <1$ is not necessary and it holds for any real value of
the parameter $\beta $. The Abel-Plana formula in its standard form (see,
for instance, \cite{Most97}) is obtained from here taking $\beta =0$.

The components $\langle T_{rr}\rangle _{\mathrm{s}}$ and $\langle T_{\rho
\rho }\rangle _{\mathrm{s}}$\ are found by the formulae similar to (\ref%
{trrn}) and (\ref{Troros1}) with the replacement ${\mathcal{D}}%
_{E}^{(k\geqslant 1)}(x,x^{\prime })\rightarrow {\mathcal{D}}_{\mathrm{sub}%
}(x,x^{\prime })$. The resulting expressions for $\langle T_{r}^{r}\rangle _{%
\mathrm{s}}$ and $\langle T_{\rho }^{\rho }\rangle _{\mathrm{s}}$
are obtained from formula (\ref{T00sGen0}) for the energy density
by the replacement $I_{\nu _{\sigma }}\left( z\right) \rightarrow
\lbrack I_{\nu _{\sigma }+n_{\sigma }}\left( z\right) +I_{\nu
_{\sigma }}\left( z\right) ]/2$ with $z=\tilde{r}^{2}/2s$ in the case of $%
\langle T_{r}^{r}\rangle _{\mathrm{s}}$ and by the replacement $I_{q|\mu
_{\delta }|}\left( z\right) \rightarrow \lbrack I_{q|\mu _{\delta }|+\delta
\mu _{\delta }/|\mu _{\delta }|}\left( z\right) +I_{q|\mu _{\delta }|}\left(
z\right) ]/2$ with $z=\rho ^{2}/2s$ in the case of $\langle T_{\rho }^{\rho
}\rangle _{\mathrm{s}}$. After summation over $\sigma $ and $\delta $ we see
that $\langle T_{r}^{r}\rangle _{\mathrm{s}}=\langle T_{\rho }^{\rho
}\rangle _{\mathrm{s}}=\langle T_{0}^{0}\rangle _{\mathrm{s}}$. Other
components are found from relations (\ref{ContEq}). It can be additionally
checked that the resulting tensor is traceless.

Now we consider the behavior of the string induced part in the VEV of the
energy density in the asymptotic regions of the parameter $y$. For large
values of this parameter, $y\gg 1$, introducing in (\ref{T00sGen}) a new
integration variable $z=yv$ and expanding the integrand over $1/y$, to the
leading order we find%
\begin{equation}
\langle T_{0}^{0}\rangle _{\mathrm{s}}\approx -\frac{(\tilde{r}/\rho
)^{2/\alpha -2}}{4^{2+1/\alpha }\pi ^{3}\alpha ^{2}\rho ^{6}}%
\int_{0}^{\infty }du\,G_{q,\gamma }(u)\frac{\sinh (u\pi )|\Gamma (1/\alpha
+2+iu)|^{2}}{\Gamma (1/\alpha +1/2)\Gamma (1/\alpha +5/2)},  \label{largey}
\end{equation}%
for $\rho \gg \tilde{r}$. In particular, for $\alpha <1$ the expectation
value induced by the string and magnetic flux vanishes on the core of the
global monopole. From formula (\ref{largey}) it follows that at large
distances from the string, the string induced energy density is suppressed
by the factor $(\tilde{r}/\rho )^{2/\alpha -2}$ with respect to the
corresponding quantity in the absence of the global monopole.

As the vacuum expectation value given by (\ref{T00sGen}) diverges on the
string core corresponding to $y=0$, in the limit $y\ll 1$ the main
contribution into the sum over $l$ comes from large values $l$ and we can
use the uniform asymptotic expansion for the functions $I_{l/\alpha \pm
1/2}(v)$. As the next step, we replace the summation over $l$ by the
integration. After some intermediate calculations, for the vacuum average of
the energy density induced by the presence of the string and magnetic flux
to the leading order one finds%
\begin{equation}
\langle T_{0}^{0}\rangle _{\mathrm{s}}\approx -\frac{1}{60\pi ^{3}\rho ^{6}}%
\int_{0}^{\infty }du\,u(u^{2}+1)(u^{2}+4)G_{q,\gamma }(u).  \label{Smally}
\end{equation}%
The expression on the right-hand side of this formula does not depend on the
parameter $\alpha $ and coincides with the corresponding quantity in the
geometry of a cosmic string with magnetic flux when the global monopole is
absent ($\alpha =1$, see below). This means that near the cosmic string the
most relevant contribution to the vacuum expectation value comes from the
string itself.

For small values of the parameter $\alpha $, corresponding to strong
gravitational field, the order of the modified Bessel functions in (\ref%
{T00sGen}) is large. Again, replacing these functions by the corresponding
uniform asymptotic expansion we can estimate the integral over $v$ by the
Laplace method. In this way it can be seen that the main contribution comes
from $l=1$ term and one finds%
\begin{equation}
\langle T_{0}^{0}\rangle _{\mathrm{s}}\approx -\frac{\exp [-(1/\alpha )\ln
(1+2y+2\sqrt{y(1+y)})]}{16\pi ^{3}\tilde{r}^{6}\alpha ^{3}y^{3/2}(1+y)}%
\int_{0}^{\infty }du\sinh (u\pi )G_{q,\gamma }(u).  \label{Smallalpha}
\end{equation}%
As we see, in this limit the vacuum expectation values are exponentially
suppressed. The similar feature takes place for the VEV of the
energy-momentum tensor and the fermionic condensate induced in the global
monopole bulk by the presence of boundaries on which the fermionic field
obeys the MIT bag boundary condition \cite{Saha04}.

In the case when the global monopole is absent, $\alpha =1$, the summation
over $l$ in the expression (\ref{T00sGen}) for $\langle T_{0}^{0}\rangle _{%
\mathrm{s}}$ can be done by using the formula \cite{Prud} $%
\sum_{l=0}^{\infty }(2l+1)I_{l+1/2}(x)=e^{x}\sqrt{2x/\pi }$. Redefining a
new variable $z=yv$ we have
\begin{equation}
\langle T_{0}^{0}\rangle _{\mathrm{s}}=-\frac{1}{8\pi ^{3}\rho ^{6}}%
\int_{0}^{\infty }dz\ z^{2}e^{-z}F_{q,\gamma }(z)\ .
\end{equation}%
By taking into account the expression (\ref{Fz}) for $F_{q,\gamma }(z)$, the
integral over $z$ is evaluated explicitly using the formula from \cite{Prud}%
. After some intermediate steps, we arrive to the following expression:%
\begin{equation}
\langle T_{0}^{0}\rangle _{\mathrm{s}}=-\frac{1}{60\pi ^{3}\rho ^{6}}%
\int_{0}^{\infty }du\,u(u^{2}+1)(u^{2}+4)G_{q,\gamma }(u).
\end{equation}%
Note that by making use of the formula
\begin{eqnarray}
A_{m}(q,\beta ) &=&\int_{0}^{\infty }du\,u^{2m-1}\left[ \sum_{\lambda =\pm 1}%
\frac{1}{e^{2\pi (u/q+i\lambda \beta )}-1}-\frac{2}{e^{2\pi u}-1}\right]  \nonumber \\
&=&\frac{(-1)^{m-1}}{2m}\left[ q^{2m}B_{2m}(\beta )-B_{2m}\right] ,
\end{eqnarray}%
where $B_{2m}(\beta )$ and $B_{2m}$ are the Bernoulli polynomials and
Bernoulli numbers respectively, the energy density is also presented in the
form%
\begin{equation}
\langle T_{0}^{0}\rangle _{\mathrm{s}}=-\frac{1}{60\pi ^{3}\rho ^{6}}%
\sum_{\delta =\pm 1}\left[ A_{3}(q,|\beta _{\delta }|)+5A_{2}(q,|\beta
_{\delta }|)+4A_{1}(q,|\beta _{\delta }|)\right] .
\end{equation}%
For the case $(q-1)/(2q)>\gamma $ the explicit calculation by this formula
gives the result
\begin{eqnarray}
\langle T_{0}^{0}\rangle _{\mathrm{s}} &=&-\frac{\pi ^{-3}\rho ^{-6}}{241920}%
[367+189q^{2}(-1+12\gamma ^{2})-21q^{4}(7-120\gamma ^{2}+240\gamma ^{4})
\notag \\
&&+q^{6}(-31+588\gamma ^{2}-1680\gamma ^{4}+1344\gamma ^{6})].
\label{T00String}
\end{eqnarray}%
In the presence of both chiralities the result given by (\ref{T00String})
should be doubled and it coincides with the formula derived in \cite{Mello4}%
. Considering the special case given by (\ref{gamSp}), from (\ref{T00String}%
) we obtain formula (\ref{TooStr}). Hence, though we have derived formula (%
\ref{TooStr}) for integer values of $q$, we see that it is valid for
non-integer $q$ as well.

\section{Conclusions}

\label{conc}

In this paper we investigate quantum vacuum effects for a massless fermionic
field induced by a composite topological defect in a six dimensional
spacetime. The spacetime is a direct product of the two dimensional cosmic
string and three dimensional global monopole geometries. The corresponding
heat kernel is constructed and on the base of this, the Green function is
evaluated for a field with positive chirality. The consideration of a
negative chirality field is similar. As the corresponding geometry with a
global monopole and in the absence of a cosmic string was investigated
previously in \cite{Mello1}, here we mainly concentrate on the effects
induced by the cosmic string and magnetic flux along the core of the string.
With this aim, the complete Green function is presented as the sum of two
terms. The first one contains information only on the global monopole defect
and the second one is induced by the presence of the cosmic string and
magnetic flux. For points away from the string the second term in the Green
function is finite in the coincidence limit and can be directly used for the
evaluation of the vacuum expectation value of the energy-momentum tensor.

In section \ref{sec:special} we have considered a special case when the
parameters of the cosmic string and magnetic flux are connected by relation (%
\ref{gamSp}) and $q$ is an integer. In this case the summation over $n$ in
the mode-sum for the heat kernel can be done explicitly and the Euclidean
Green function is presented in the form (\ref{Desp}). The vacuum expectation
value of the energy-momentum tensor is constructed from the Feynman
propagator by making use of formula (\ref{EM}). The energy density induced
by the string and magnetic flux is given by formula (\ref{tt}). The
corresponding vacuum stresses along $r$- and $\rho $-directions coincide
with energy density. The other components of the vacuum stress are found
from the covariant continuity equation for the energy-momentum tensor. We
have explicitly checked that the part of the energy-momentum tensor induced
by the string and flux is traceless.

The Green function and the vacuum expectation value of the energy-momentum
tensor for the general case of the string and flux parameters are discussed
in section \ref{sec:Gen}. By using the Abel-Plana summation formula, the
corresponding subtracted Green function is presented in the form (\ref%
{DsubGen}) where the parts giving the divergences in the
coincidence limit are explicitly cancelled out. Further, this
Green function is used for the evaluation of the corresponding
energy-momentum tensor. The vacuum energy density is given by
formula (\ref{T00sGen}) and, as for the special case, the stresses
along $r$- and $\rho $-directions coincide with energy density.
The remained components are found from relations (\ref{ContEq}).
As for pure cosmic string geometry, the VEV of the energy-momentum
tensor depends only on the fractional part of the magnetic flux
and is an even function of this parameter. We have investigated
the vacuum energy-momentum tensor in
asymptotic regions of the parameters. At large distances from the string, $%
\rho \gg \tilde{r}$, the leading term of the asymptotic expansion of the
energy density is given by formula (\ref{largey}). In this case the string
induced energy density is suppressed by the factor $(\tilde{r}/\rho
)^{2/\alpha -2}$ with respect to the corresponding quantity in the absence
of the global monopole. For points near the string core, $\rho \ll \tilde{r}$%
, to the leading order the string induced part in the vacuum average
coincides with the corresponding quantity in the geometry of a cosmic string
with magnetic flux when the global monopole is absent. For small values of
the parameter $\alpha $, corresponding to strong gravitational fields, the
vacuum expectation values are exponentially suppressed.

\section*{Acknowledgment}

E.R.B.M. thanks Conselho Nacional de Desenvolvimento Cient\'{\i}fico e Tecnol%
\'{o}gico (CNPq) for partial financial support, FAPESQ-PB/CNPq (PRONEX) and
FAPES-ES/CNPq (PRONEX). A.A.S. was supported by the Armenian Ministry of
Education and Science Grant No. 119 and by Conselho Nacional de
Desenvolvimento Cient\'{\i}fico e Tecnol\'{o}gico (CNPq).

\end{document}